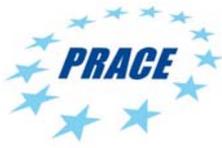

**Partnership for Advanced Computing in Europe**

# Scalability of the Plasma Physics Code GEM

B. Scott[a*], V. Weinberg[b†], O. Hoenen[a], A. Karmakar[b] , L. Fazendeiro[c]

[a]*Max-Planck-Institut für Plasmaphysik IPP, 85748 Garching b. München, Germany*
[b]*Leibniz Rechenzentrum der Bayerischen Akademie der Wissenschaften, 85748 Garching b. München, Germany*
[c]*Chalmers University of Technology, 412 96 Gothenburg, Sweden*

**Abstract**

We discuss a detailed weak scaling analysis of GEM, a 3D MPI-parallelised gyrofluid code used in theoretical plasma physics at the Max Planck Institute of Plasma Physics, IPP at Garching near Munich, Germany. Within a PRACE Preparatory Access Project various versions of the code have been analysed on the HPC systems SuperMUC at LRZ and JUQUEEN at Jülich Supercomputing Centre (JSC) to improve the parallel scalability of the application. The diagnostic tool Scalasca has been used to filter out suboptimal routines. The code uses the electromagnetic gyrofluid model which is a superset of magnetohydrodynamic and drift-Alfvén microturbulance and also includes several relevant kinetic processes. GEM can be used with different geometries depending on the targeted use case, and has been proven to show good scalability when the computational domain is distributed amongst two dimensions. Such a distribution allows grids with sufficient size to describe conventional tokamak devices. In order to enable simulation of very large tokamaks (such as the next generation nuclear fusion device ITER in Cadarache, France) the third dimension has been parallelised and weak scaling has been achieved for significantly larger grids.

## 1. Introduction

The code GEM addresses electromagnetic turbulence in tokamak plasmas [1,2,3,4,5]. Its main focus is on the edge layer in which several poorly understood phenomena are observed in experiments. A feature of medium- to large-tokamak plasmas is that the edge layer is described by a pedestal feature in the plasma pressure profile, mostly due to the temperature but also the particle density: for a few cm inside the plasma boundary the gradient is very steep: R/L of order 50, where L is the scale length and R is the toroidal major radius (see Fig. 1 for a sketch of a tokamak). The ratio R/L controls the toroidal drive effects involving the pressure gradient upon the turbulence. Inside of this edge layer, the profile is more gradual with R/L in the range of 5 to 10. Experimental findings are that if the pedestal top temperature is higher, so will the core value be, and hence in a fusion reactor the yield should be better. Understanding this is a central theme in magnetically confined plasma research. Direct simulation with realistic parameters with appropriate scale separation is one of the avenues. For a reactor tokamak the edge-pedestal layer can be several hundred plasma skin depths or ion gyro radii wide, and this defines the spatial grid to be used in computations. Fortunately, the grid must have this resolution in only the two dimensions perpendicular to the magnetic field, but this field is sheared in a tokamak, so the coordinate system must be aligned with it to get this benefit. Additionally, the underlying equations involve elliptic-equation solves for the field variables which are spatially inhomogeneous, so that sparse-matrix techniques are required. Mapping the tokamak field-based coordinate system such that its grid cells are mostly conformal is work we had done previously. The version of GEM which does this at present calculates fluctuations on a given background and is called GEMZA. A future version which treats proper stratification is under development but will have the same computational infrastructure. Hence, our project has been involved with improving the scalability of GEMZA in preparation for complete edge-pedestal simulations for large tokamaks by its successor.

GEMZA has been run up to now on conventional tokamak cases for which ASDEX-Upgrade at the Max-Planck-IPP in Garching, Germany, is an example. A grid to cover the edge layer only is $64 \times 4096 \times 16$, where the last number is the number of planes along the toroidal magnetic field, and powers of two are used to facilitate the multigrid (MG) field solver. With the radial coordinate being the conformal flux radius, the radial extent is about the last 8 percent of the minor radius, and the 4096 is for the poloidal angle direction. The $64 \times 4096 \times 16$ count gives the number of grid points in what we call the x- and y- and s-directions, respectively. For the purposes of the project, this defines the thin-strip, small-tokamak case. For the small tokamak case we also use a medium strip (doubles the 64 to 128) which combines the edge layer with the scrape-off layer, the region encompassing the first scale-length drop of the profiles outside of the closed field line region. We also use a thick-strip case which includes the outer core region, essentially doubling the thickness again such that we have a grid of $256 \times 4096 \times 16$ for the small tokamak case.

Due to the magnetic fluctuation level, which is what eventually decides how many planes are needed, a jump to a larger system requires doubling all three spatial coordinate grid cell counts, not just two as in the electrostatic computations which are more common. So the thin-strip, medium-tokamak case is $128 \times 8192 \times 32$, and the next one for large tokamaks is $256 \times 16384 \times 64$. These are the most important base cases to test. Eventual tokamak pedestal simulation will require a thicker strip, to include the open field line region ("scrape-off-layer", SOL), leading to the medium-strip cases, and possibly also the outer-core region, leading to the thick-strip cases. The largest case we foresee is the thick strip version of the large-tokamak case, whose grid is $1024 \times 16384 \times 64$. The small- medium- and large-tokamak cases correspond roughly to ASDEX-Upgrade, the JET tokamak in Culham, UK, and the ITER reactor-plasma experiment under construction in Cadarache, France. The relevance of the large-tokamak case is the aim to directly simulate the ITER edge layer in order to predict the performance of the pedestal, i.e., the temperature at the pedestal top which, due to the tendency of the core to have a threshold value of R/L, controls the central temperature and therefore due to the scaling of fusion reaction rates the performance of the entire tokamak.

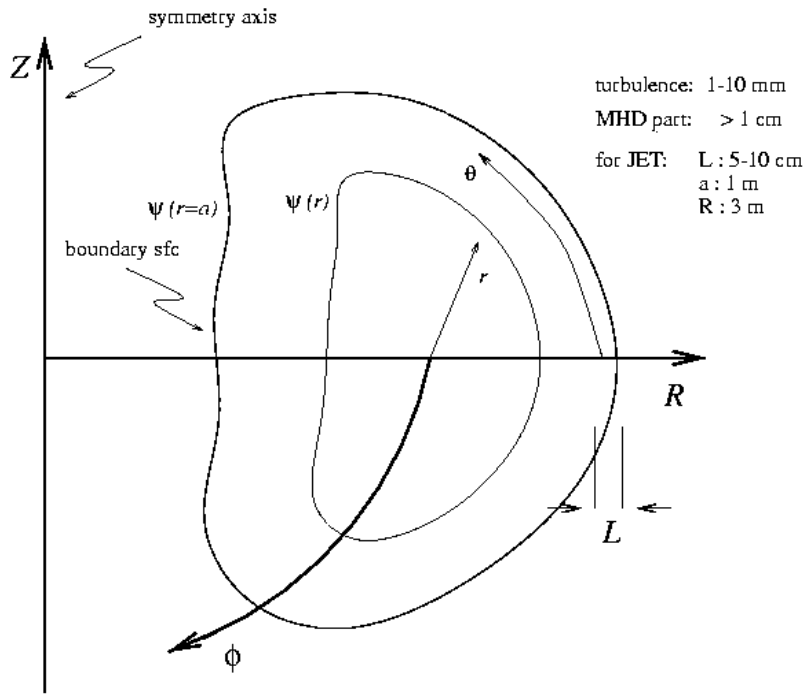

Figure 1. Sketch of a tokamak. The contours shown are rotated about the symmetry axis to make figures of revolution. The magnetic field lies in these flux surfaces, representing isosurfaces of the magnetic flux variable. The boundary surface defines r=a so that a is the minor radius and the value of R at the geometric centre defines the major radius. A thin shell of width L denotes the edge layer, and defines the thin-strip case mentioned in the text. Typical scales for the JET tokamak are given. The dimensions of the ITER tokamak will be one factor of two scaled up from JET. The coordinates in GEMZA are prescribed functions x=x(r), s=ϕ, and y=y(θ,ϕ) such that its gradient is perpendicular to the magnetic field which is mostly toroidal but also has a poloidal component.

Our interest in weak scaling of the code to larger cases is therefore evident. Work up to now has been on the smallest case and successfully exhibited hard scaling up to 512 cores on the HPCFF platform. Following this base case, our goal was to establish weak scalability such that a problem with $64 \times 128 \times 1$ grid points per core could be scaled up to large-tokamak cases within the same or similar wall-clock time. The reason for the "1" in the s-direction dimension is that the field solver works in the xy-plane and does not involve the s-direction. This project's investigations essentially confirmed the solver's role as the principal bottleneck in the basic time step. These nine cases, the 3x3 matrix of small- medium- large-tokamak and thin- medium- thick-strip, going from 512 to 131072 cores in steps of a factor of 2, form the basis of the project. The essential result of the project is the weak scaling information given in Fig. 3, but the work of the project was a matter of how to get there.

## 2. Description of the code versions

The code solves the equations representing gyrofluid field theory for tokamak plasmas. The basic model is a gyrokinetic description [6,7] which is then simplified to a gyrofluid description [1,6]. In both cases the distribution function or density, temperature, and other such moment variables follow gyrocentres (the centres of gyromotion around the strong magnetic field) rather than particles so that high frequencies are filtered out of the dynamics. All such codes have the same basic structure. A set of equations for the moment variables is evolved during one time step. The character of these equations is advection along characteristics, which for this model is better described by a series of Poisson brackets in each of a number of two-dimensional (2-D) planes within a 4-D dynamical phase space, or a 3-D space in the fluid case. This involves only nearest-neighbour exchanges and hence simple *MPI_Sendrecv* calls. Weak scaling of this should be straightforward. At the end of the time step, a set of field equations is solved for field potentials (in this case, two) which are determined by the newly-evolved moment variables. However, these field equations are elliptic equations of the Helmholtz type, $(a - \text{div } b \text{ grad}) p = S$ for a field variable $p$ given known scalar quantities $a$, $b$, and $S$. The essential property of elliptic equations is that the field variable $p$ at one grid point is determined by the source variable $S$ at all grid points globally. Generally, sparse-matrix techniques are required, and these are well known to be harder to scale. In our case, besides I/O, the difficulty in scaling is presented by the solver for these Helmholtz equations. In principle the solver is 30 to 50 percent of the effort. But this is enough such that failure of the solver to scale causes failure of the overall code to scale.

Various versions of the MPI-parallelised code were set up for the detailed scaling analysis. Since the main difficulty of the code is the solver, we have focused on its improvement. For most of the study the I/O routines were eliminated to isolate the solver features. As a control case, one version of the code was set up with a dummy solver, denoted (dummy), basically replacing the Helmholtz equations with scalar ones of the type $A p = S$ with $A$ and $S$ known scalars. This served as a control case against which the scaling of all other versions was compared. Then, a conjugate gradient (CG) version of the solver was used. Here, the main communication is nearest-neighbour but it also includes a dot product which involves the *MPI_Allreduce* function with the *MPI_SUM* attribute. The method of choice is actually multigrid (MG). Most of the cases involve parallelisation in all three dimensions, i.e., also the x-direction which is the thin dimension across the strip. Multigrid involves a hierarchy of grids, from the main one at the finest level to factor-of-two coarsening all the way down to two points across the thin dimension globally (for anisotropic domain size such as this, the coarsening stops when the shortest-direction limit is reached, since the eigenvalue spread of the Helmholtz operator determines convergence [8], and the spread is controlled by this direction). For this the code has used V-cycles in the past, and we call this version (MGV). A variant used in the computational literature is to use what are called U-cycles, in which the coarsening only proceeds to the level at which two points exist on each core in the x-direction. We call this version (MGU). See Fig. 2 for a comparison of the V- and U-cycles schemes. These four versions were tested without the I/O routines. Finally, a fifth version of MGU was tested with the I/O routines put back in, representing the code as we would use it for production cases. The reason MGU was the choice for this was that it performed the best in the sans-I/O testing.

For the CG algorithm see the text by Golub and Ortega [9] and for the MG algorithm using V-cycles see the tutorial by Briggs [8]. For the method using U-cycles see the paper by Xie and Scott [10].

A simple list of the versions we tested is given in the following items:

- **Dummy version**
  The dummy version uses a simple scalar formula for the field variables in terms of their sources, using no boundary or sum information. The only MPI communication is the boundary exchange in the main time step. This version also serves as a basic test of purely nearest-neighbour scalability.

- **CG version**
  The CG version uses a conjugate-gradient method for the solver iteration with no preconditioner. Originally developed as an alternative to multigrid schemes for large systems, but whose algorithm scales worse than O(N) for large grids.

- **MGU version**
  The MGU version (Multigrid with U-cycle scheme), which has less communication but slower, and parallelism-dependent, convergence. This version has been developed within the project.

- **MGV version**
  The code's original solver version, from 2008. Convergence is independent of grid size, but communication cost is increased when the x-direction is parallelised with less than one active grid point per core.

- **MGV version with I/O**
  The code's original version, from 2008. I/O had been improved previously but was still using serial I/O through one core. We did not test with this due to the superior scalability of the MGU version.

- **MGU version with I/O**
  After the MGU version showed the best scalability, we restored the I/O routines in this one, and developed the many-files version within the project. Each plane in the s-direction writes in parallel to a separate file. The s-direction is the most separated, "most distant in bandwidth", so that architectures with only sparse I/O can still use this version.

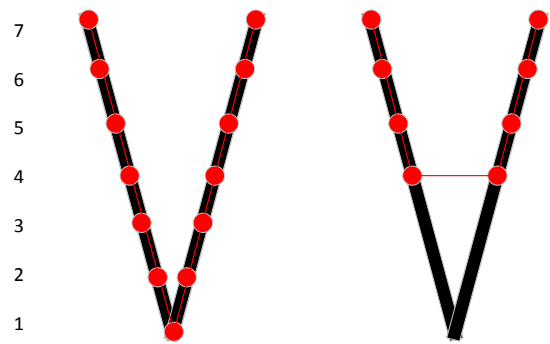

Figure 2. Sketch of the V-cycle (left) and U-cycle (right) convergence schemes in the multigrid method for a sparse matrix solver. The number giving the level indicates the base-2 log of the number of grid points. In the V-cycle coarsening proceeds by factors of two until only two grid points are left in the shortest direction. If that direction is parallelised there can be too few points on a core for acceptable communication cost. The U-cycle scheme stops when there are two points left per core. Convergence degrades slightly but scalability is better. The U-cycle method was formulated and analysed by [10].

**3. Scaling results**

Scaling measurements have been done on the HPC systems SuperMUC (IBM System x iDataPlex) at LRZ and JUQUEEN (IBM Blue Gene/Q) at Jülich Supercomputing Centre (JSC). The Scalasca utility [11] has been used to analyse the runtime behaviour of the code and break the total measured wall-clock time down into the time spent within MPI ("MPI"), pure user functions ("USR") and functions calling subprograms or MPI ("COM"). The Scalasca analysis has been mainly done on the thin node islands of the SuperMUC system. Each thin node island consists of 512 SandyBridge-EP Intel Xeon nodes, and each node is equipped with 16 cores clocked with max. 2.7 GHz. The peak performance on SuperMUC is 345.6 GFlops/node. We present scaling results within one island (max. 8192 cores) and using completely filled two (16384 cores) and four (32768 cores) islands on SuperMUC. Four (out of 18) thin node islands is the maximum allocatable number of islands on SuperMUC for regular projects. On JUQUEEN the scaling analysis has been done on up to 131072 cores. Each IBM PowerPC A2 node of JUQUEEN has 16 cores clocked with 1.6 GHz. The peak performance on JUQUEEN is 204.8 GFlops/node.

Fig. 3 compares the overall runtime of the MGU version on both platforms. In all cases the test run use case was for 1000 time steps on JUQUEEN and 5000 on SuperMUC, which coincidentally found a speed ratio between SuperMUC and JUQUEEN of about 5. Running only one task per core (as recommended for SuperMUC), the "effective" peak performance on PowerPC A2 is 202 GFlops/node. Considering also the different clock speed of the cores this accounts for a factor of approx. 3.5 in comparison with SuperMUC. The remaining factor of 1.5 is often found in many applications not tuned for in-order architectures like PowerPC A2. Intel Xeon is an out-of-order architecture. This helps masking processor stalls due to dependencies among instructions.
The wall clock time for all subsequent runs was below 900 sec, except for the failure cases (CG on the larger grids showed convergence degradation on JUQUEEN and either crashed or failed to finish within 1800 sec on SuperMUC on more than 8192 cores, and the original I/O version showed increasing cost above 4096 cores on JUQUEEN).

Figs. 4-6 show the results of the Scalasca scaling analysis on SuperMUC: the overall wall-clock time (Fig. 4 (a)), the total time spent within MPI calls (Fig. 4 (b)), time spent in the dominating MPI functions *MPI_Allreduce* (Fig 5(a)) and *MPI_Sendrecv* (Fig. 5 (b)) and time spent in non-MPI regions of the code (Fig. 6). Mind that time spent in the "MPI" (Fig. 4 (b)) + "USR" (Fig. 6 (a)) + "COM" (Fig. 6 (b)) regions of the code equals the total wall-clock time (Fig. 4 (a)). The instrumentation with Scalasca did not have any significant influence on the runtime behaviour and the execution time of the program. Fig. 6 recovers the expectation that the non-MPI portions of the code scale as they should since no communication is being performed. The weak scaling curve is flat. Comparing with the results for the MPI routines one can see that these routines are a significant fraction of the code's execution. The scalability of the MPI routines it uses therefore determines the scalability of the entire code. This is very significant to the CG solver due to its dependence through the vector dot product in the algorithm upon the *MPI_Allreduce* function.

In the following we discuss the scaling results of the various versions of the code in more detail

- **Dummy version**
  The dummy version showed very low cost in total (Fig. 4 (a)) and an *MPI_Sendrecv* cost that scaled very well to larger systems even though at finite cost (Fig. 5 (b)). This essentially eliminated the first main concern about the code, that communication in more dimensions might be worse than the same amount in one dimension. The weak scaling of the dummy version was used as a baseline for the other versions of the code. On SuperMUC the dummy version wall-clock time was 212.0 and 237.2 and 305.1 sec for the thin-strip cases of 512 and 4096 and 32768 cores, respectively (Fig. 4 (a)). On JUQUEEN the times were 245.6 and 250.5 and 274.7 sec, respectively. In light of the multigrid experience (see below) this suggests s-direction communication as the first bottleneck.

- **CG version**
  The Scalasca analysis isolated in the CG case the *MPI_Allreduce* operations as the most costly (Fig. 5 (a)). This is due to the vector dot product in the CG method which involves every grid point in the 2D plane. Figs. 4 and 5 (a) document the limits of scalability. Even the algorithm doesn't scale, due to a particular feature of the physics in the code. We first tried CG on a 2D fluid case and it looked promising indeed, using only 10 iterations per time step, converging across time steps. However, this depends on the 2D fluid physics of slower time scales for larger space scales. In the physics of tokamak edge turbulence, all the nonlinear dynamics is underlain by drift wave turbulence, which is a competition between eddy dynamics in 2D, as in a fluid model, but now against electromagnetic wave dynamics in the 3rd dimension (the s-direction). This latter is just as fast for larger as for smaller perpendicular scales. As a result, we find a need for 500 iterations for the smallest case, rapidly increasing with case size, easily indicated by the code's time-step diagnostics. Weak scaling of the CG version of the code therefore degraded clearly by 4096 cores and rapidly collapsed after that.

- **MGU version**
  In both MG versions the 512 core case does not communicate in the x-direction through MPI since there is only the one core, while for all other cases there are at least two cores. On the 512 core case with MGU the time was 420 sec. For larger cases the x-direction communication incurs a penalty, but only in the step to 1024 cores is it a scalability issue. The times for all the larger cases were in the 700-800 sec range. The increase is due to the solver since the dummy version, acting as a control, shows no such penalty (Fig. 4a). However, once this is paid, the scaling on subsequent cases is such that the time rises from 682.8 sec for 1024 cores to 878.0 on 32768 cores. Degradation is measured as a time ratio from case to case (Fig. 3 (b)). Though there is some noise on the trend, the log of the average ratio above 1024 cores is about 0.05 and no obvious increase is seen above 8192 cores which is one island on the SuperMUC architecture. On JUQUEEN this result was even better. The 512 core case had a time of 429.4 sec and the others ranged from about 720 to about 790 with a noise level of about 15 sec. The log of the time ratio averaged 0.013, with a negative value in some cases, indicative of the noise. From 1024 to 131072 cores the efficiency was 720.5/786.9 = 0.92 with an error bar of about 0.03. For our purposes, this result defined the success of the project.

- **MGV version**
  Originally it was feared that the observed failure in scalability above 4096 cores prior to this project was due to the solver. This might have been true on capacity cluster systems, but the removal of the I/O routines has shown that the main bottleneck is the unimproved I/O. The results on both SuperMUC and JUQUEEN and the Scalasca analysis on SuperMUC have found that the MGV is only marginally inferior to the MGU version, although as Fig. 4 shows the penalty going from MGU to MGV is visible. Fig. 4b shows this is due to the MPI routines, as expected. This rather large penalty of between 10 and 20 percent from circa 2048 to 16384 cores is entirely due to the main difference: the degradation caused by communication when the number of coarse grid points is less than one per core in the x-direction.

- **MGU version with I/O**
  With the successful test on the MGU version of the code, the I/O was put back in. From previous experience we already knew that for a single file blocked I/O (each processor appends a file in turn, or sends to PE 0 which does the write) was much better than striped I/O (write the file as if only one core wrote a variable on the whole grid, one variable at a time). We found the I/O cost on JUQUEEN for the blocked single-file I/O was negligible up to 4096 cores but caused a penalty of about 30 percent on 32768 cores. Even though the jump in the penalty on SuperMUC was less severe (Fig. 4 (a)), a new method is needed for the largest cases. We tried multi-file I/O, in which the blocked single-file I/O is used separately for each plane in the s-direction. For the small- medium- and large-tokamak cases 16, 32 and 64 files are written. Unfortunately we ran out of time before we could get the 131072 core case but the others show a cost increase below 15 percent going from 4096 to 65536 cores. While this penalty is acceptable, we expect scaling with this method to break down for larger cases, so a search for an I/O solution is underway subsequent to this project.

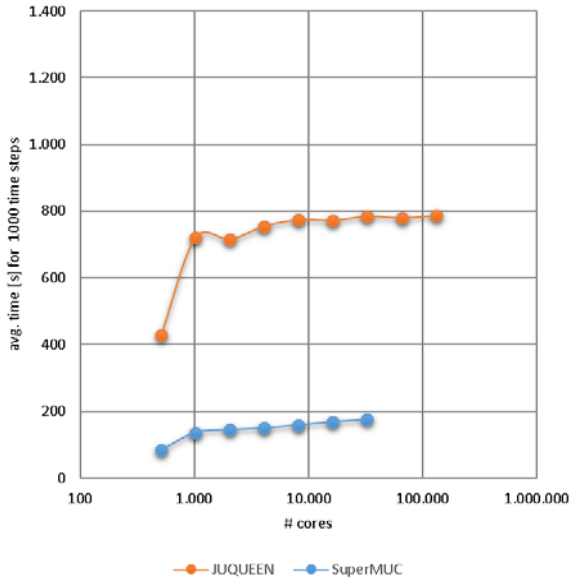
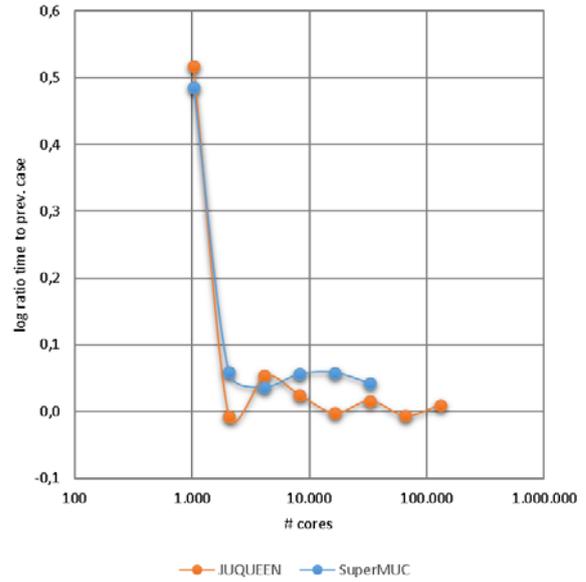

**(a)**            **(b)**

Figure 3. Weak scaling of the MGU version of GEMZA on SuperMUC (MUC) and JUQUEEN (JUQ). Fig. (a) shows the wall-clock time as a number of the MPI tasks (=cores) for 1000 time steps, showing the speed factor of about 5 between the two platforms. Fig. (b) presents the log of the time ratio for each case compared to the previous case (0.0 is perfect). The jump from 512 to 1024 cores is the one-time penalty for increasing the number of cores in the x-direction above unity. Weak scaling is slightly better on JUQUEEN, with an average log time ratio of 0.0126 compared to 0.05 on SuperMUC.

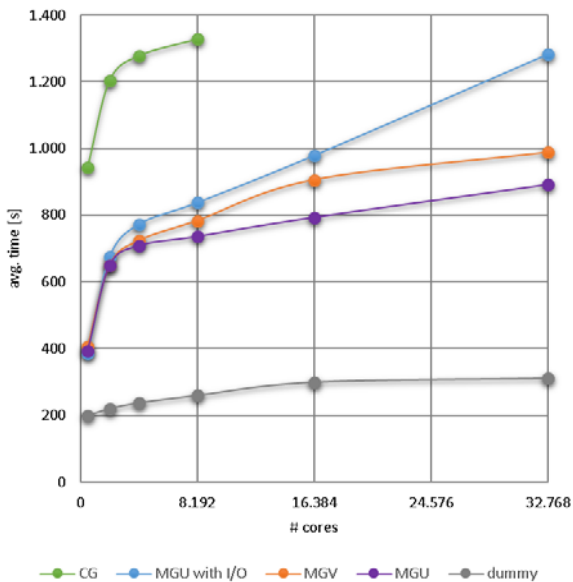
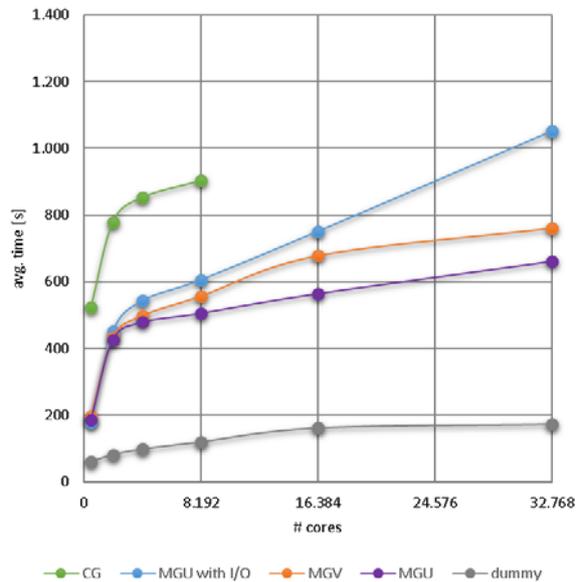

**(a)**            **(b)**

Figure 4. Weak scaling of various versions of GEMZA on SuperMUC. Fig. (a) shows the overall time, the same information as in Fig. 3 (a). Fig. (b) presents the portion spent in the MPI routines, from the Scalasca analysis. CG is the conjugate-gradient version, dummy is the version without the solver, and the others are multigrid versions with U- (MGU) or V-cycles (MGV) iteration schemes, and the one case with I/O shows the increased cost of I/O for the MGU version. I/O cost increased worse than this above 4k cores on JUQUEEN. The increased cost of the CG version was made worse for larger cases due to convergence degradation.

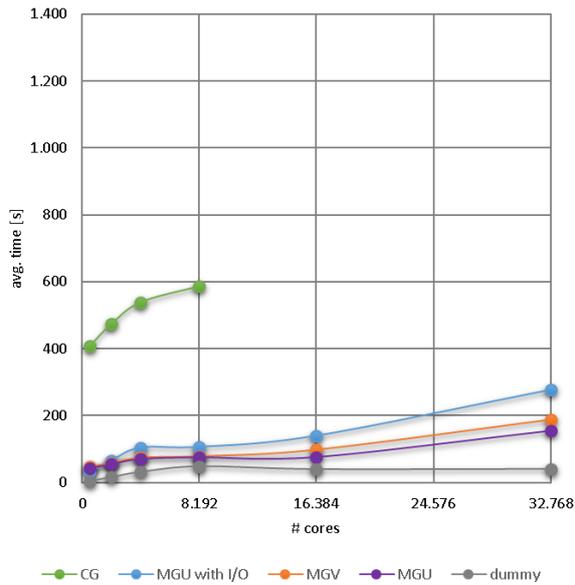
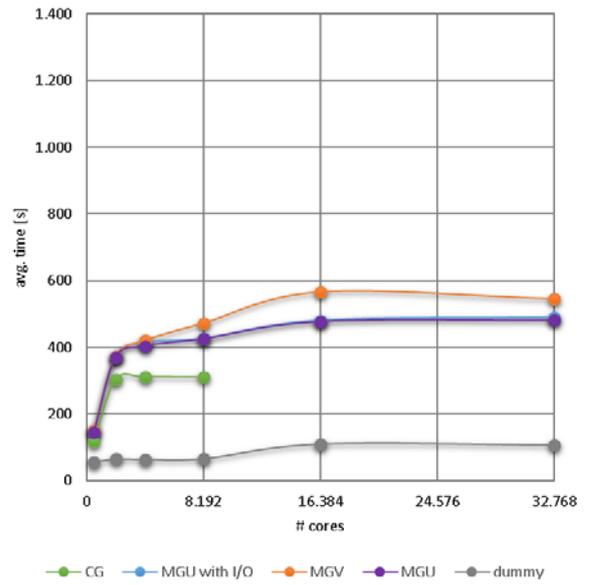

**(a)**　　　　　　　　　　　　　　　　　**(b)**

Figure 5. Weak scaling of the dominating MPI routines for the various versions of GEMZA on SuperMUC from the Scalasca analysis, showing *MPI_Allreduce* in Fig. (a) and *MPI_Sendrecv* in (b). The *MPI_Allreduce* cost in the CG version, due to the dot product in the CG scheme, exceeded all the boundary exchanges in the multigrid versions, even the V-cycles version MGV with less than one grid point per core on the coarsest levels.

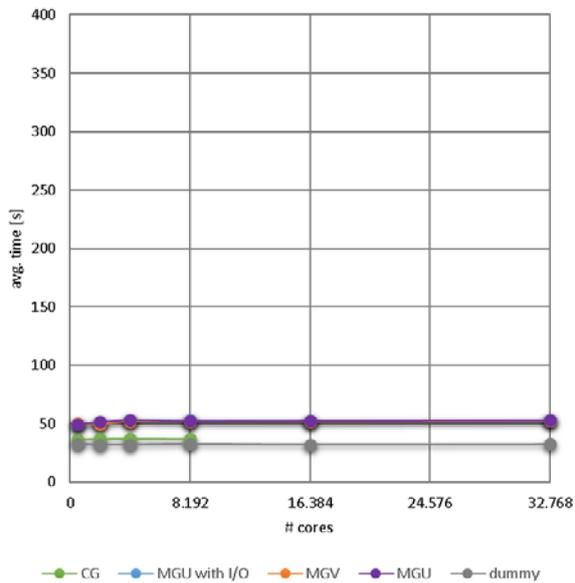
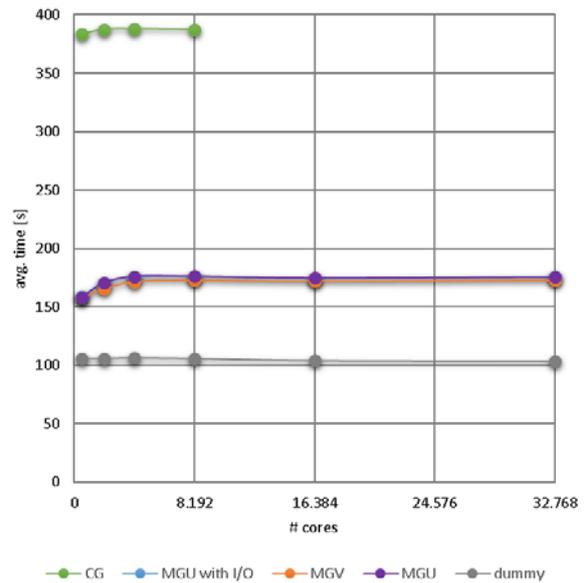

**(a)**　　　　　　　　　　　　　　　　　**(b)**

Figure 6. Weak scaling of the non-MPI routines of the code for the various versions of GEMZA on SuperMUC from the Scalasca analysis. Fig. (a) displays the results for "USR"-functions not calling any subroutines and (b) the "COM"-functions which call subroutines or MPI calls, however the time spent within the "USR" or "MPI" regions is not counted by Scalasca. Mind that the scale of the y-axis is a factor of 3.5 smaller than in the other figures.

## 4. Summary and Outlook

The main goal of the project was to establish weak scalability of the gyrofluid code GEMZA to larger systems: if we can afford a thin-strip, small-tokamak case $64 \times 4096 \times 16$ on 512 cores then we would like to run the largest planned thick-strip, large-tokamak case $1024 \times 16384 \times 64$ on 131072 cores in the same or similar wall clock time. Main loop wall clock time was the yardstick to test several cases, each of which having $64 \times 128 \times 1$ grid points per core in x, y, s. The final result is that a penalty of about 30 percent was paid once, when the number of cores in the x-direction changed from 1 to 2. For larger cases, the weak-scalability goal was well reached. On a machine like SuperMUC the standard production runs which go for 400k time steps can be run within 18 hours (this was done in Oct 2013 on Hydra, a platform with similar architecture to SuperMUC). On a BG/Q system like JUQUEEN the code was 5 times slower but the scalability was better. The 131k-core case would run on a BG/Q in about 100 wall-clock hours, were sufficient time made available. This was the definition of success when the project was proposed.

Our main conclusion is that the goal of running larger systems commensurate with the ITER tokamak edge-pedestal layer region with similar wall-clock time with the same number of grid points per core as standard cases was achieved (in principle, since actual runs require production resources). The diagnostic tool Scalasca greatly helped filtering out suboptimal options. With the help of Scalasca a version of the code has been found for which acceptable weak scaling is confirmed. The relevance of the results obtained is essentially the demonstration of feasibility of planned large-system runs. B. Scott is extending the physics content of the code to treat real stratification. For this purpose new equations have been derived [12]. The improvement of the I/O scheme using systems such as the ADIOS library [13] and possible extension to a hybrid MPI/OpenMP scheme are being explored. This new version of the code will have the same structure and operation set as the current one and is planned to be used in a future PRACE project.

## Acknowledgements

This work was financially supported by the PRACE project funded in part by the EUs 7th Framework Programme (FP7/2007-2013) under grant agreement nos. RI-283493 and RI-312763. The results were obtained within the PRACE Preparatory Access Type C Project 2010PA1505 "Scalability of gyrofluid components within a multi-scale framework".